# Transient quantum isolation and critical behavior in the magnetization dynamics of half-metallic manganites

T. Pincelli[1,2], R. Cucini[1], A. Verna[3], F. Borgatti[4], M. Oura[5], K. Tamasaku[5], H. Osawa[6], T.-L. Lee[7], C. Schlueter[7], S. Günther[8], C.H. Back[8,9], M. Dell'Angela[1], R. Ciprian[1§], P. Orgiani[1,10], A. Petrov[1], F. Sirotti[11], V. A. Dediu[4], I. Bergenti[4], P. Graziosi[4], F. Miletto Granozio[12], Y.Tanaka[13], M. Taguchi[5,14,15], H. Daimon[14], J. Fujii[1], G. Rossi[1,2], and G. Panaccione[1*]

[1] Istituto Officina dei Materiali (IOM)-CNR, Laboratorio TASC, in Area Science Park, S.S.14, Km 163.5, I-34149 Trieste, Italy
[2] Dipartimento di Fisica, Università di Milano, Via Celoria 16, I-20133 Milano – Italy
[3] Dipartimento di Scienze and CNISM, Università degli Studi Roma Tre, Via della Vasca Navale 84, I-00146 Roma, Italy
[4] Consiglio Nazionale delle Ricerche – Istituto per lo Studio dei Materiali Nanostrutturati (CNR-ISMN), via P. Gobetti 101, I-40129 Bologna, Italy
[5] RIKEN SPring-8 Center, Kouto 1-1-1, Sayo-cho, Sayo-gun, Hyogo 679-5148, Japan
[6] Japan Synchrotron Radiation Research Institute (JASRI), Kouto 1-1-1, Sayo-cho, Sayo-gun, Hyogo 679-5198, Japan
[7] Diamond Light Source, Harwell Science and Innovation Campus, Didcot OX11 0DE, United Kingdom
[8] Institut für Experimentelle Physik, Universitat Regensburg, D-93040 Regensburg, Germany
[9] Department of Physics, Technical University Munich, D-85748 Garching b. München, Germany
[10] CNR-SPIN, UOS Salerno, 84084 Fisciano, Italy
[11] Laboratoire Physique de la Matière Condensée (PMC), Ècole Polytechnique, 91128 Palaiseau Cedex, France
[12] CNR-SPIN, UOS Napoli, Complesso Universitario di Monte Sant'Angelo, Via Cinthia, I-80126 Naples, Italy
[13] Graduate School of Material Science, University of Hyogo, 3-2-1 Kouto, Kamigori-cho, Ako-gun, Hyogo 678-1297, Japan
[14] Nara Institute of Science and Technology (NAIST), 8916-5 Takayama, Ikoma, Nara 630-0192, Japan
[15] Toshiba Nanoanalysis Corporation, Kawasaki, Kanagawa 212-8583, Japan
§Deceased

*Corresponding author: panaccioneg@elettra.eu

## ABSTRACT

We combine time resolved pump-probe Magneto-Optical Kerr Effect and Photoelectron Spectroscopy experiments supported by theoretical analysis to determine the relaxation dynamics of delocalized electrons in half-metallic ferromagnetic manganite $La_{1-x}Sr_xMnO_3$. We observe that the half-metallic character of $La_{1-x}Sr_xMnO_3$ determines the timescale of both the electronic phase transition and the quenching of magnetization, revealing a quantum isolation of the spin system in double exchange ferromagnets extending up to hundreds of picoseconds. We demonstrate the use of time-resolved hard X-ray photoelectron spectroscopy (TR-HAXPES) as a unique tool to single out the evolution of strongly correlated electronic states across a second-order phase transition in a complex material.

## I. INTRODUCTION

The description of the competition between delocalization and confinement of electrons in a solid is a fascinating challenge in condensed matter physics [1, 2], and it is deeply connected with the electronic correlation effects that determine the properties of strongly correlated materials.

Electron-electron interactions beyond the Hartree-Fock exchange give rise to numerous exotic behaviors such as, among others, high temperature superconductivity, half-metallicity, metal-insulator transitions, multiferroicity and spin-charge separation [1-5, 69]. This is realized in condensed matter when partially filled -$d$ or -$f$ bands are narrowed to the point at which the mean field "Fermi sea" approximation is no longer valid.

In perovskite transition metal oxides (TMO), the band originating from the $3d$ orbital of the transition metal is split by the octahedral symmetry of the oxygen ligands in a $t_{2g}$ band, and in a narrow $e_g$ band. In presence of strong exchange interaction, next-to-nearest neighbor interactions between cations are determined by superexchange mechanisms, i.e. coupling through the non-magnetic ligand anion. Depending on the specific details of the electronic structure, superexchange can favor ferromagnetic or antiferromagnetic alignment, and tend to localize or delocalize carriers [2, 70].

In manganites as $La_{1-x}Sr_xMnO_3$ (LSMO) electronic and transport properties are described by the superexchange interaction called double exchange (DE) [3-5]. The mobility of carriers is coupled to itinerant magnetism in a delicate form: the energy of the crystal is lowered by delocalizing the majority $e_g$ electrons between Mn sites determining metallicity [7, 8]. This is the driving mechanism that favors spin alignment, as Hund's rule coupling produces parallel spin alignment between $e_g$ and localised $t_{2g}$ electrons, resulting in ferromagnetism with a net separation between spin-up ($e_{g\uparrow}$, $t_{2g\uparrow}$) and spin-down ($e_{g\downarrow}$, $t_{2g\downarrow}$) bands, i.e. half-metallic behavior [6, 9-11].

Wide-band-gap half-metals are ferromagnets of special interest: half-metallicity was described by de Groot and coworkers [6], for materials having a metallic electronic Density of States (DOS) in one spin channel but a gap in the DOS for the other spin state, allowing 100% spin polarized electrons near the Fermi level, a highly desirable property in spintronics materials.

In a DE system, therefore, the second-order transition from paramagnetic to magnetically ordered phase is accompanied by an electronic phase transition, as the density of states is increased close to the Fermi level owing to the delocalization of carriers [2-5]. A new perspective in the investigation of complex, strongly correlated systems is offered by the realization of studies in the time domain, i.e. with stroboscopic methods after ultrashort optical excitation. This method allows to disentangle relevant excitations in the solid: ultrafast optically-induced dynamics creates a 'window of observation' by producing the dynamical separation of spin, lattice and charge thermodynamic baths due to the different timescales of their mutual interactions [12].

In ferromagnetic *3d* metals a 'femto-magnetism' regime (<1ps) has been proposed where competition between order parameters follows the relative out-of-equilibrium timescales [13]. Conversely, half-metals present a 'bottlenecked' magnetization dynamics, as the phonon assisted spin-flip scattering mechanism described by Eliott and Yafet is suppressed [14-18], and the demagnetization time can be three orders of magnitude longer in comparison to a classic *3d* ferromagnet [17].

Although a number of studies have been reported on the ultrafast dynamics of half-metallic systems [19-21], a clear understanding of the interplay between localized and delocalized electrons in the magnetization dynamics of DE systems is lacking and a direct measurement of the electronic properties in the transient states with both chemical and magnetic sensitivity has been rarely achieved.

To address these issues, we have performed pump-probe experiments on a prototypical half-metal: the optimally Strontium-doped Lanthanum Manganite $La_{0.66}Sr_{0.33}MnO_3$, for which the relatively large bandwidth suppresses polaronic self-trapping [3], making this system the closest to the ideal DE ferromagnet. Here we combine two complementary bulk-sensitive pump-probe techniques: Time-Resolved Magneto-Optical Kerr effect (TR-MOKE) and Time-Resolved HArd X-ray core level PhotoElectron Spectroscopy (TR-HAXPES).

These two techniques have been used to provide a twofold perspective on the dynamical behaviour of this material. With TR-MOKE, we detect the evolution of the macroscopic magnetization, arising from localized $t_{2g}$ momenta. With TR-HAXPES, we observe the specific evolution of the localization/delocalization of the $e_g$ states. Combining optical methods relating to macroscopic observables with spectroscopic information sensitive to the details of the electronic structure, proves to be a reliable method to approach correlated materials. A complete separation of the relevant timescales of spin and charge systems is observed, and the magnetization dynamics in LSMO can be described as a 'transient quantum isolation' of the spin system with respect to the charge degree of freedom. The characteristic critical time for the collapse of long-range magnetic order is obtained by means of TR-MOKE and TR-HAXPES reveals that delocalized electrons play an active role in the dynamics of the magnetization on a timescale of 100 ps and above.

## II. METHODS

### A. EXPERIMENTAL DETAILS

Time resolved and temperature dependent HAXPES experiments were performed at BL19LXU at SPring-8, Japan, equipped with a SCIENTA R4000-10kV electron energy analyser at grazing incidence geometry (<4° angle between the X-ray beam and the surface plane) and a spot size of $40 \times 500$ μm$^2$ on the sample. The overall energy resolution (analyser + beamline) was kept below 500 meV at the selected photon energy of 7.94 keV. At this energy, the inelastic mean free path of 2p electrons in LSMO is $8.6 \pm 0.8$ nm, corresponding to a mean escape depth of $7.9 \pm 0.8$ nm [41].

A Ti:Sapphire oscillator and a regenerative amplifier optical laser system were used for the pump-probe setup, with a pulse energy of 0.6-4.0 mJ/pulse, a wavelength of 800 nm, a duration of 100 fs (FWHM) and a repetition rate of 1 kHz. For pump-probe HAXPES experiments the so-called H-mode of operation was used, consisting of a single electron bunch (pulse width 50 ps) and a continuous bunch train, *e.g.*, 1 bunch + 11/29-filling. Single bunch of H-mode filling was selected by using an x-ray chopper with a repetition rate of 1 kHz, with jitter error well below the pulse duration (5ps). The timing between Synchrotron light and pump laser pulse was monitored by a fast photodiode mounted at the bottom of sample manipulator.

MOKE experiments were performed, in the longitudinal MOKE configuration, with an 800 nm pump and a 400 nm probe, to avoid optical bleaching effects and to maximise the Kerr angle sensitivity. Two methods have been used: for the highest accuracy, a balanced detection scheme was employed, using a Wollaston prism; for rapid acquisition the crossed-polarizers method was employed. All measurements were performed at remanence, reversing the magnetization at every delay. The ellipsometric analysis was performed measuring both ellipticity and rotation, and evaluating the full complex Kerr angle.

In both experiments, the same $La_{0.67}Sr_{0.33}MnO_3$ sample was used. It was a 100 unit cells thick LSMO film grown by reactive molecular beam epitaxy on $(LaAlO_3)_{0.3}(Sr_2TaAlO_6)_{0.7}$ (LSAT), a transparent, insulating perovskitic oxide. LSAT has a lattice mismatch below 0.3% with LSMO, and the film can be therefore considered strain-relaxed throughout the thickness. In all pump-probe experiments, the absorption length of the 800 nm pump is 40 nm, equal to the thickness of the LSMO film, which can be considered uniformly excited. For more details about the experimental setup and characterization of samples, see Supplementary Materials [22].

**B. CALCULATION DETAILS**

Modelling the electronic structure of strongly correlated oxides requires methods beyond the classical Hartree-Fock single Slater determinant. For the analysis of satellites in photoemission, the configuration-interaction (CI) method is generally used, where the ground state is obtained as a linear combination of N-electron wave-functions including low energy excited configurations. The problem is then solved by direct diagonalization on this basis. To reconstruct the photoemission process, the CI approach is combined with the Anderson impurity model to treat the core-hole produced by photoemission. Given the short lifetime of the core-hole the inclusion of a single Mn atom and the surrounding oxygen octahedron is sufficient to describe photoemission spectra of LSMO. To reconstruct Mn 2p lines,, the Mn 2p orbital, the Mn 3d and the O 2p have to be included.

Experimental observation of additional peaks appearing in bulk sensitive photoemission spectra has required the introduction of a further, highly hybridized state [42] called the coherent state, which has recently been demonstrated to arise from non-local (between different Mn sites) charge transfer processes [10].

We used three configurations as basis states: $3d^4$, $3d^5\underline{L}$, and $3d^5\underline{C}$. The last two representing charge transfer excitations from the ligand 2p states and from the coherent state. In each configuration, intra-atomic multiplet, spin-orbit effect, hybridization between Mn 3d and O 2p states, on-site coulomb repulsion and core-hole attractive potential are accounted for.

The spectrum of MCD-HAXPES is given by calculating the helicity dependent spectral function. In the calculations for XPS with right and left circular polarized x-ray in Fig. 2, an unpolarized component is also included in order to take into account that the degree of polarization of the X-ray is not 100% and the direction of the incident X-ray is not completely parallel to z-direction.

For more details about the specific choice of parameters for the calculation, see the Supplementary Materials Sect. 9.

### III. RESULTS

The results of optical experiments are shown in Fig.1-2, tracing the time evolution of the macroscopic magnetisation of LSMO as a function of the pump-probe delay at various initial sample temperatures. As depicted in Fig 1a, the system is excited with an 800 nm pump pulse of 70 fs duration and 2 mJ/cm$^2$ fluence; the probe is of the same duration, but has a wavelength of 400 nm. The dynamical reflectivity curves (Fig. 1c), obtained concurrently with the TR-MOKE measurement, show first a sharp increase, and subsequently a decay to a first equilibrium value within 1.2 ps after the pump pulse (inset of Fig. 1c). The equilibration time results of 160 ± 70 fs as obtained by exponential fit between the peak edge and the flattening at 1.2 ps. This rapid decay can be considered, in metallic samples, as a signature of the thermalization of electrons with the lattice, pointing to a hot-carrier character of the process [23].

In the topmost curve of Fig. 1c, after the first 10 ps, a slower increase of the reflectivity takes place and reaches its maximum at about 200 ps, followed by a gradual decrease of the reflectivity as the sample cools down at even longer times. The observed long timescale of the reflectivity dynamics is a peculiar trait of DE manganites (including LSMO [24,25]), and involves a process called Dynamical Spectral Weight Transfer (DSWT) representative of systems crossing a quantum critical phase under optical excitation [25-27].

The comparison of reflectivity curves below and above the Curie temperature $T_c$ ($T_c$ = 335 K in present case) reveals the presence of DSWT, namely the pump-induced increase in the reflectivity extending over long timescales, and disappearing above Tc. As the DE ferromagnet is pushed towards a higher resistivity phase with non collinear core spin momenta, spectral weight is removed from the Drude band in the optical spectrum and higher energy transitions are enhanced [28], inducing a variation of the refractive index and accordingly of the reflectivity [24]. Reflectivity results shown in Fig. 1c clearly identify a rearrangement of the electronic structure in LSMO unfolding over several hundreds of ps, thus strongly influencing the mobility of electrons close to Fermi energy over a long timescale, simultaneously with the demagnetisation process.

In the presence of such a complex rearrangement of the electronic structure, indeed, the dynamics of the refraction and absorption index make the connection between MOKE signal and magnetization less straightforward [29]. We applied two color, doubly modulated MOKE with Kerr angle modulus reconstruction (see Supplementary Materials Sect. 2), to reduce the effect of the rearrangement of spectral weight in the optical spectrum. The signal obtained in this way describes reliably the long timescale dynamics, in the 10 ps – 1 ns range.

In Fig.1d, one observes a minimum at a delay of around 200 ps in the curve measured at 150 K, corresponding to a transient equilibrium between the magnetization and demagnetization. For larger pump-probe delays, heat dissipation allows the sample to cool down and the curve takes an upward turn pointing back to the maximum magnetization. The curve measured at 300 K displays not only slower dynamics but also a shift in the position of the minimum that moves outside the accessible range of delay time. Such behavior, termed critical slowing-down, reflects the increase of characteristic time of the magnetic fluctuations in proximity of a thermodynamic phase transition [33-36, 38, 40]. It is a universal scaling law and applies to a vast range of critical processes, including the present case of a second-order ferromagnetic to paramagnetic transition, and results in a continuous divergence following a negative power law.

Temperature-dependent ellipticity measurements mirroring the behavior of the magnetization (see Supplementary Materials Sect. 2) have been performed (Inset Fig. 2), to monitor the long time-scale collapse of the magnetization over a wide range of temperature. The characteristic time $\tau_{sl}$ is retrieved for each temperature by fitting an exponential function to the time evolution plotted in the inset of Fig. 2. The resulting $\tau_{sl}$ is plotted vs. sample initial temperature in Fig. 2, which shows that the studied temperature range can be divided into a critical region (305 K – 345 K) and a non-critical one (165 K – 305 K). In the non-critical region we observe a slow quench of the magnetisation, ranging from $74 \pm 2\ ps$ at 150 K to $186 \pm 3\ ps$ at 300 K, values compatible with a wide gap half-metallic systems [16].

After the rapid increase in temperature, the system starts to cool by heat transfer through the substrate. Given that the samples are thin films on insulating ceramics (LSAT), the heat transport is so slow and so weakly temperature dependent that can represented by a temperature-independent recovery rate, not affecting the fitting of the decay constants.

In previous studies the timescales are found to be above the nanoseconds, with no observable temperature dependence [67, 68]. We confirmed this by solving the heat equation in a simple model of the LSMO/LSAT heterostructure, obtaining negligible variations of the film cooling dynamics with different initial temperatures (see Supplementary Materials Sect. 8).

In the vicinity of $T_c$ the demagnetization process is slowed down by the divergence of the spin specific heat. This can be modeled with the methods employed to analyze critical behavior in proximity of a phase transition. In the critical region, we can estimate the characteristic time of the divergence, applying the dynamical scaling theory [38, 39], with

$$\tau \propto |T - T_C|^{-zv}$$

where $v$ is the correlation length exponent following the definition that spatial correlation $\propto |T - T_C|^{-v}$, and z is the dynamical exponent. A power-law fit to the curve in Fig. 2 in the range 305 K – 330 K gives $zv = 1.28 \pm 0.16$, in reasonable agreement with $zv = 1.35 \pm 0.01$ found theoretically for the 3D double exchange ferromagnet, and in very good agreement with the result of models accounting for long range order $zv = 1.251$ [40,22] (see also Supplementary Materials Sect. 3).

This confirms that the critical slowing-down observed at higher temperatures arises from the divergence of the spin specific heat and suggests that the critical behavior of LSMO is well described by the model of a three dimensional ferromagnet in which DE is the dominant interaction.

We also measured the demagnetization dynamics at different fluences, and, as expected in a second-order phase transition, we did not find a critical fluence (see Supplementary Materials Sect. 6). In a half-metallic material, indeed, the magnetic system has extremely long characteristic timescales. With respect to these, the electronic and lattice system appear to immediately equilibrate and to produce a sudden change in temperature, which shifts the system along the M(T) curve. The temperature variation and the final temperature to which the magnetic system is driven are the only parameters relevant to the magnetic dynamics (see Supplementary Materials Sect. 7).

In the non-critical region in Fig. 2 (165 K <T <305 K), the intrinsic timescale is often attributed to the anisotropy coupling of spin-lattice interaction [16-18]. The gap in the minority states that characterizes a half-metallic material, indeed, provokes a strong suppression of any type of scattering mechanism involving a spin flip in the final states, as it would require sufficient energy to reach the minority states above the gap. The strong coupling through phonon-assisted Elliott-Yafet scattering is therefore suppressed, as is direct electron-magnon interaction. Only second order interactions involving spin-flips in a virtual state can couple with the spin system: inelastic spin-phonon coupling [18] and second-order correlated electron-magnon interactions [39].

In the former picture, the lattice vibrations couple, via spin-orbit interaction, to magnetic anisotropy and excite the localized spin system. A rough estimate for $\tau_{sl}$ is customarily obtained using the time-energy uncertainty relationship and the typical energies of the interaction involved. Note, however, that neglecting the quantum mechanical details of the scattering process only allows to define an approximate lower boundary for $\tau_{sl}$.

In our case, considering an anisotropy energy of 5 ÷ 1 μeV [37], $\tau_{sl}$ is estimated in the range of 800 ps ÷ 4 ns for $T = 150 \div 300\ K$. The observed increase of $\tau_{sl}$ vs. temperature is qualitatively in agreement with the temperature dependence of the anisotropy constants [37]. Yet, the overall tendency to overestimate this value implies that other processes are simultaneously at play.

Of particular interest is, instead, the latter interaction mechanism. Indeed, it has been demonstrated that superposition of spin-up electron excitations and virtual magnons (also called "spin-polaron processes") can produce the lowest energy excitations for minority spins, allowing demagnetization on long timescales. These results indicate that, for temperatures below the critical region, the demagnetization characteristic time(s) is described by strongly suppressed electron-spin and spin-lattice relaxation mechanisms.

As the magnetization dynamics detected by MOKE owes its largest contribution to the localized $t_{2g}$ magnetic momenta, we conclude that it describes a magnetic system close to a Heisenberg ferromagnet, which undergoes transient quantum isolation: no obvious single-particle high-energy process can couple the excitation of the magnetic system with charge excitations, giving rise to a long timescale demagnetization.

We now address the question whether or not the connection of $t_{2g}$ ferromagnetic alignment with the emergent mobility of the $e_g$ carriers described by the DE Hamiltonian in Eq. 1 breaks down at these timescales. In order to probe the dynamics of the specific electronic structure, we performed chemically and magnetically sensitive pump-probe Hard X-ray Photoemission Spectroscopy (HAXPES) experiments. HAXPES guarantees bulk sensitivity (up to an integration depth of 10 nm [41]), removing uncertainties arising from any surface effects [42-45]. Thus, the HAXPES results can be directly compared with those of MOKE in the present study.

The Mn 2p core-level spectrum of LSMO, as measured at equilibrium by HAXPES, is shown in Fig. 3a, where a distinct satellite is present at a binding energy around 640 eV. Such low binding energy satellites have been observed in HAXPES spectra for many oxide systems: they are spectroscopic fingerprints of the strength of electron delocalization and thus of the metallicity of the oxides [42, 45-49].

The features of the Mn *2p* spectrum can be disentangled by the theoretical analysis shown in Fig. 3b. The calculated total Mn *2p* spectrum with $Mn^{3+}$ valence is obtained by including three electronic configurations describing the final states of Mn ions, namely $2p^53d^4$, $2p^53d^5\underline{L}$ and $2p^53d^5\underline{C}$, corresponding respectively to the ground state $3d^4$, the charge transfer from a ligand hole $3d^5\underline{L}$ and the charge transfer from hybridized states close to the Fermi energy, $3d^5\underline{C}$. The $3d^5\underline{C}$ component has a significant spectral weight at the energies corresponding to the well-screened satellites.

Spectra are calculated with a V* value for hybridization of 1.17 eV, corresponding to the best agreement with experimental curves [45]. The main contribution to the low binding energy structures arises from the initial $3d^5C$ configuration. Note, however, that with V* being non-zero, both $3d^4$ and $3d^5\underline{L}$ initial states can project intensity onto the $3d^5\underline{C}$ final state, yet their calculated contribution to the satellite intensity is negligible [50,51].

To investigate the connection between the well-screened features and the long-range magnetization, we measure the polarization and temperature dependence of the Mn 2p core level for LSMO, in analogy to investigations performed on other complex materials [52-55]. Fig. 4 shows the comparison between the experimental and calculated magnetic circular dichroism in HAXPES of the Mn 2p core level. According to the photoemission selection rules, photons with -1(+1) helicity, i.e., right (left) circularly polarized x-rays, create selectively spin-polarized core-holes [56]. When the valence electrons involved in the screening mechanism are also spin-polarized with a spin parallel to the core-hole spin, a large magnetic dichroism occurs in the photoemission as the result of exchange interaction between the core and valence states.

In Fig. 4 the Mn 2p satellites of LSMO show strong magnetic circular dichroism. The regions around 639.5 eV and 651.5 eV of binding energy (BE) reflect a well-defined final state with parallel spins in the core and valence states, displaying the expected opposite dichroism for spin-orbit split $2p_{1/2}$ and $2p_{3/2}$ components [57,58]. Both the intensity of the Mn *2p* satellites and the magnitude of the dichroism decrease with an increasing temperature (inset in Fig. 3a and differences at T = 200 and 300 K in Fig. 4), thus reflecting the thermal reduction of the magnetization (see also Supplementary Materials Sect. 5) [45]. Model calculations reproduce accurately the dichroic lineshapes [56, 59] when hybridization is finite (solid curves in Fig. 4), allowing to pin the strongly spin polarized character to the $3d^5\underline{C}$ contribution.

The dichroic and temperature-dependent analysis of the Mn *2p* line-shape, backed by a solid theoretical description of the complex substructures of this doublet allows to establish the low binding energy satellites as a precise probe of the hybridization and spin state of delocalized carriers responsible of the DE mechanism.

The pump-probe Mn $2p_{3/2}$ HAXPES results are presented in Fig. 5b as a function of pump-probe delays (from -900 ps to +3.1 ns), with a laser-off spectrum included for comparison (see also Supplementary Materials Sect. 4). Experimental geometry is shown in Fig. 5a. In panel 5b one notices that the spectra measured at negative delays are identical to the laser-off spectrum and, independent of the delay time, no extra broadening or energy shifts are present: this excludes the presence of space charge [60-62] and/or heat pile-up artifacts. The coincidence of pump and probe pulses (zero delay), has been determined self-consistently within the measurement as the first delay at which a line-shape evolution is observed.

A significant reduction of the well-screened satellite peak intensity is clearly observed within the first 100 ps after a pump pulse, followed by a slow recovery at later times. Intensity variations are also observed in the Mn *2p* manifold centered at 644 eV of binding energy, as highlighted by Fig. 5c. The increase in spectral weight at high binding energies is observed concomitantly to the decrease in the low binding energy satellite also for quasi-static temperature variations. However, the difference at 0 ps delay, while displaying a large decrease at low binding energies, does not show an enhanced high binding energy manifold. This, supported by the optical evidence gathered previously, suggests that the electronic structure evolves on timescales longer than the pulse duration (50 ps), and only after 100 ps a new transient equilibrium is formed at higher temperatures.

A summary of the pump-probe time dependence of the well screened satellite is presented in Fig. 6, where the relative area difference between the laser-off and laser-on spectra in the energy range 638.5 - 640.5 eV is shown for different delays. The top panels in Fig. 6 show the corresponding Mn $2p_{3/2}$ lineshapes at three delays, while the fits of the satellite peaks highlight the intensity variation arising from the delocalized carriers.

The amplitude and timescale of the evolution observed in Fig. 6 are comparable to the ones observed in low temperature (150 K) TR-MOKE in Fig. 1. This, combined with the observation of non-equilibrium lineshapes at short delays with a long 50 ps probe pulse (see Fig. 5c), confirms that the dynamics of the delocalized carriers takes place with the same timescale of the macroscopic magnetic momentum, indicating an unbroken, coherent evolution of the DE electronic structure.

We thus can connect the complementary measurement techniques in a coherent picture. By optical infrared pumping the electronic system is excited at short timescales and the hot electron distribution relaxes very rapidly (few fs to 1 ps) by transferring heat to the lattice. The magnetic relaxation, due to excitation of localized momenta, is dominated by weak mechanisms, owing to the system half-metallicity. Such slow magnetic dynamics keeps the DE mechanism active for long timescales, thus preventing the triggering of a sharp metal-insulator transition. This is in contrast to what is observed in other TMOs, where the metal-insulator transition can be triggered even in absence of a contemporary variation of the associated order parameter [63, 64].

In summary, we have performed combined pump-probe optical and photoelectron experiments to study the magnetization dynamics in double exchange half-metallic LSMO. The ensemble of our results is consistent with the collapse of the electronic hybridization over an extended timescale, pointing to the transient quantum isolation of the spin system. The DE mechanism tightly connects electronic delocalization and magnetic interaction in LSMO, it therefore explains the slow dynamics of the half-metal, whose spectral features show out of equilibrium configurations for delays up to hundreds of picoseconds after optical excitation.

These results establish TR-HAXPES, in combination with support techniques, as a powerful technique to explore the dynamics of specific features of the electronic structure in complex materials. The upcoming development of brilliant sources with high temporal resolution and high repetition rate even in the hard x-ray range will allow to explore the dynamics at the timescales of electron-electron interactions and thus shed a new light on the fascinating problems arising from electronic correlation [65, 66].

## ACKNOWLEDGEMENTS

This work has been partly performed in the framework of the nanoscience foundry and fine analysis facility (NFFA-Italy), funded by Ministero dell'Istruzione, dell'Università e della Ricerca (MIUR). The experiments at BL19LXU of SPring-8 were carried out with the approval of JASRI (proposal nos. 2015B1162, 2016A1289, and 2017A1323). The authors are grateful to the members of the engineering team of the RIKEN SPring-8 Center for their technical assistance during the installation of the HAXPES apparatus. Thanks are due to F. Cilento, G. van der Laan, W. Hübner, F. Grasselli, A. Regoutz and D.J. Payne for fruitful discussion. A. V. acknowledges partial financial support from the PRIN project NEWLI (code 2015CL3APH) of the MIUR.

**References**

1. N.F. Mott, and R. Peierls, *Discussion of the paper by de Boer and Verwey*. Proc. Phys. Soc. **49,** 72 (1937).
2. M. Imada, A. Fujimori and Y. Tokura, *Metal-insulator transitions*. Rev. Mod. Phys. **70,** 1039–1263 (1998).
3. M.B. Salamon and M. Jaime, *The physics of manganites: Structure and transport*. Rev. Mod. Phys. **73,** 583–628 (2001).
4. E. Dagotto, T. Hotta and A. Moreo, *Colossal magnetoresistant materials: the key role of phase separation*. Physics Reports **344,** 1–153 (2001).
5. J. M. D. Coey, M. Viret and S. von Molnár, *Mixed-valence manganites*. Advances in Physics **48,** 167–293 (1999).
6. R.A. de Groot, F.M. Mueller, P.G. van Engen and K.H. J. Buschow, *New Class of Materials: Half-Metallic Ferromagnets* Phys. Rev. Lett. **50,** 2024–2027 (1983).
7. P.W. Anderson and H. Hasegawa. *Considerations on Double Exchange*. Phys. Rev. **100,** 675–681 (1955).
8. P.G. de Gennes, *Effects of Double Exchange in Magnetic Crystals*. Phys. Rev. **118,** 141–154 (1960).
9. M. Bowen, A. Barthélémy, M. Bibes, E. Jacquet, J.-P. Contour, A. Fert, F. Ciccacci, L. Duò, and R. Bertacco *Spin-Polarized Tunneling Spectroscopy in Tunnel Junctions with Half-Metallic Electrodes*. Phys. Rev. Lett. **95**, 137203 (2005).

10. A. Hariki, A. Yamanaka and T. Uozumi, *Orbital- and spin-order sensitive nonlocal screening in Mn 2p X-ray photoemission of La 1−x Sr x MnO3*. EPL **114,** 27003 (2016).
11. M. I. Katsnelson, V.I. Irkhin, L. Chioncel, A. Lichtenstein, and R.A. de Groot *Half-metallic ferromagnets: From band structure to many-body effects*. Rev. Mod. Phys. **80,** 315–378 (2008).
12. A. Kirilyuk, A. Kimel and T. Rasing, *Ultrafast optical manipulation of magnetic order.* Rev. Mod. Phys. **82,** 2731–2784 (2010).
13. P. Tengdin, W. You, C. Chen, X. Shi, D. Zusin, Y. Zhang, C. Gentry, A. Blonsky, M. Keller,, P. M. Oppeneer, H. C. Kapteyn, Z. Tao and M. M. Murnane , *Critical behavior within 20 fs drives the out-of-equilibrium laser-induced magnetic phase transition in nickel,* Science Advances, 4, 9744 (2018).
14. R.J. Elliott, *Theory of the Effect of Spin-Orbit Coupling on Magnetic Resonance in Some Semiconductors.* Phys. Rev. **96,** 266–279 (1954).
15. Y. Yafet, Y. *g Factors and Spin-Lattice Relaxation of Conduction Electrons,* Solid State Physics (eds. Seitz, F. & Turnbull, D.) **14,** 1–98 (Academic Press, 1963).
16. G. M. Müller, J. Walowski, M. Djordjevic, G.-X. Miao, A. Gupta, A. V. Ramos, K. Gehrke, V. Moshnyaga, K. Samwer, J. Schmalhorst, A. Thomas, A. Hütten, G. Reiss, J. S. Moodera and M. Münzenberg, *Spin polarization in half-metals probed by femtosecond spin excitation*. Nat Mater **8,** 56–61 (2009).
17. A. Mann, J. Walowski, M. Münzenberg, S. Maat, M. J. Carey, J. R. Childress, C. Mewes, D. Ebke, V. Drewello, G. Reiss, and A. Thomas, *Insights into Ultrafast Demagnetization in Pseudogap Half-Metals,* Phys. Rev. X **2,** 041008 (2012).
18. W. Hübner and K.H. Bennemann, *Simple theory for spin-lattice relaxation in metallic rare-earth ferromagnets* Phys. Rev. B**53,** 3422–3427 (1996).
19. H. Ehrke, R. I. Tobey, S. Wall, S. A. Cavill, M. Först, V. Khanna, Th. Garl, N. Stojanovic, D. Prabhakaran, A. T. Boothroyd, M. Gensch, A. Mirone, P. Reutler, A. Revcolevschi, S. S. Dhesi, and A. Cavalleri, *Photoinduced Melting of Antiferromagnetic Order in La$_{0.5}$Sr$_{1.5}$MnO$_4$ Measured Using Ultrafast Resonant Soft X-Ray Diffraction*, Phys. Rev. Lett. **106**, 217401 (2011).
20. P. Beaud, A. Caviezel, S. O. Mariager, L. Rettig, G. Ingold, C. Dornes, S-W. Huang, J. A. Johnson, M. Radovic, T. Huber, T. Kubacka, A. Ferrer, H. T. Lemke, M. Chollet, D. Zhu, J. M. Glownia, M. Sikorski, A. Robert, H. Wadati, M. Nakamura, M. Kawasaki, Y. Tokura, S. L. Johnson and U. Staub *A time-dependent order parameter for ultrafast photoinduced phase transitions* Nat. Mater. **13,** 923–927 (2014).
21. J. Li, W.-G. Yin, L. Wu, P. Zhu, T. Konstantinova, J. Tao, J. Yang, S.-W. Cheong, F. Carbone, J. A. Misewich, J. P Hill, X. Wang, R. J. Cava and Y. Zhu, *Dichotomy in ultrafast atomic dynamics as direct evidence of polaron formation in manganites,* NPJ Quantum Materials **1,** 16026 (2016).
22. See Supplemental Material for description of growth procedures and transport characterization, analysis of the kerr results, details of data analysis for fitting critical slow down and time resolved data analysis and fitting of HAXPES data.
23. M. Bonn, D. N. Denzler, S. Funk, M. Wolf, S.-Svante Wellershoff and J. Hohlfeld, *Ultrafast electron dynamics at metal surfaces: Competition between electron-phonon coupling and hot-electron transport*. Phys. Rev. B**61,** 1101–1105 (2000).
24. A.I. Lobad, A.J. Taylor, C. Kwon, S.A. Trugman and T.R. Gosnell, *Laser induced dynamic spectral weight transfer in La$_{0.7}$Ca$_{0.3}$MnO$_3$.* Chemical Physics **251,** 227 (2000).
25. R. D. Averitt, A. I. Lobad, C. Kwon, S. A. Trugman, V. K. Thorsmølle, and A. J. Taylor *Ultrafast Conductivity Dynamics in Colossal Magnetoresistance Manganites.* Phys. Rev. Lett. **87,** 017401 (2001).
26. S. A. McGill, R. I. Miller, O. N. Torrens, A. Mamchik, I-Wei Chen, and J. M. Kikkawa *Dynamic Kerr Effect and the Spectral Weight Transfer of the Manganites*. Phys. Rev. Lett. **93,** 047402 (2004).
27. A.I. Lobad, R.D. Averitt, C. Kwon and A. J. Taylor, *Spin–lattice interaction in colossal magnetoresistance manganites*. Appl. Phys. Lett. **77,** 4025–4027 (2000).
28. Y. Okimoto, T. Katsufuji, T. Ishikawa, T. Arima and Y. Tokura, Variation of electronic structure in $_L$La$_{1-x}$Sr$_x$MnO$_3$ x=0.3 as investigated by optical conductivity spectra. *Phys. Rev. B* **55,** 4206–4214 (1997).
29. E. Carpene, F. Boschini, H. Hedayat, C. Piovera, C. Dallera. E. Puppin, M. Mansurova , M. Munzenberg,X. Zhang and A. Gupta. *Measurement of the magneto-optical response of Fe and CrO$_2$ epitaxial films by pump-probe spectroscopy: Evidence for spin-charge separation*. Phys. Rev. B**87,** 174437 (2013).

30. B. Koopmans, *Laser-Induced Magnetization Dynamics* in *Spin Dynamics in Confined Magnetic Structures II* 256–323 (Springer, Berlin, Heidelberg, 2003). doi:10.1007/3-540-46097-7_8)
31. G.P. Zhang, W. Hübner, G. Lefkidis, Y. Bai and T.F. George, *Paradigm of the time-resolved magneto-optical Kerr effect for femtosecond magnetism.* Nat Phys **5,** 499–502 (2009).
32. J-Y. Bigot, *Femtosecond magneto-optical processes in metals.* Comptes Rendus de l'Académie des Sciences - Series IV - Physics **2,** 1483–1504 (2001).
33. T. Kise, T. Ogasawara, M. Ashida, Y. Tomioka, Y. Tokura, and M. Kuwata-Gonokami, *Ultrafast Spin Dynamics and Critical Behavior in Half-Metallic Ferromagnet: $Sr_2FeMoO_6$.* Phys. Rev. Lett. **85,** 1986–1989 (2000).
34. X.J. Liu, Y. Moritomo, A. Nakamura, H. Tanaka and T. Kawai, *Critical behavior of a photodisordered spin system in doped manganite.* Phys. Rev. B **64,** 100401 (2001).
35. T. Ogasawara, M. Matsubara, Y. Tomioka, M. Kuwata-Gonokami, H. Okamoto, and Y. Tokura. *Photoinduced spin dynamics in $La_{0.6}Sr_{0.4}MnO_3$ observed by time-resolved magneto-optical Kerr spectroscopy.* Phys. Rev. B **68,** 180407 (2003).
36. T. Ogasawara, K. Ohgushi, Y. Tomioka, K. S. Takahashi, H. Okamoto, M. Kawasaki, and Y. Tokura, *General Features of Photoinduced Spin Dynamics in Ferromagnetic and Ferrimagnetic Compounds.* Phys. Rev. Lett. **94,** 087202 (2005).
37. K. Steenbeck and R. Hiergeist, *Magnetic anisotropy of ferromagnetic $La_{0.7}(Sr,Ca)_{0.3}MnO_3$ epitaxial films.* Appl. Phys. Lett. **75,** 1778–1780 (1999).
38. P.C. Hohenberg and B.I. Halperin, *Theory of dynamic critical phenomena.* Rev. Mod. Phys. **49,** 435–479 (1977).
39. S. Brener, B. Murzaliev, M. Titov, and M. I. Katsnelson. *Magnon activation by hot electrons via nonquasiparticle states.* Phys. Rev. B **95**, 220409(R) (2017).
40. R. Singh, K. Dutta, and M. K. Nandy, *Critical dynamics of a nonlocal model and critical behavior of perovskite manganites.* Phys. Rev. E **93,** 052132 (2016).
41. M. Sacchi, F. Offi, P. Torelli, A. Fondacaro, C. Spezzani, M. Cautero, G. Cautero, S. Huotari, M. Grioni, R. Delaunay, M. Fabrizioli, G. Vankó, G. Monaco, G. Paolicelli, G. Stefani and G. Panaccione, *Quantifying the effective attenuation length in high-energy photoemission experiments.* Phys. Rev. B **71,** 155117 (2005).
42. K. Horiba, M. Taguchi, A. Chainani, Y. Takata, E. Ikenaga, D. Miwa, Y. Nishino, K. Tamasaku, M. Awaji, A. Takeuchi, M. Yabashi, H. Namatame, M. Taniguchi, H. Kumigashira, M. Oshima, M. Lippmaa, M. Kawasaki, H. Koinuma, K. Kobayashi, T. Ishikawa, and S. Shin, *Nature of the Well Screened State in Hard X-Ray Mn $2p$ Core-Level Photoemission Measurements of $La_{1-x}Sr_xMnO_3$ Films.* Phys. Rev. Lett. **93,** 236401 (2004).
43. Z. Sun J. F. Douglas, A. V. Fedorov, Y.-D. Chuang, H. Zheng, J. F. Mitchell and D. S. Dessau, *A local metallic state in globally insulating $La_{1.24}Sr_{1.76}Mn_2O_7$ well above the metal–insulator transition.* Nature Physics **3,** 248–252 (2007).
44. J. W. Freeland, J. J. Kavich, K. E. Gray, L. Ozyuzer, H. Zheng, J. F. Mitchell, M. P. Warusawithana, P. Ryan, X. Zhai, R. H. Kodama and J. N. Eckstein*et al. Suppressed magnetization at the surfaces and interfaces of ferromagnetic metallic manganites.* J. Phys.: Condens. Matter **19,** 315210 (2007).
45. T. Pincelli, V. Lollobrigida, F. Borgatti, A. Regoutz, B. Gobaut, C. Schlueter, T.-L. Lee,, D.J. Payne, M. Oura, K. Tamasaku, A.Y. Petrov, P. Graziosi, F. Miletto Granozio, M. Cavallini, G. Vinai, R. Ciprian, C.H. Back, G. Rossi, M. Taguchi, H. Daimon, G. van der Laan and G. Panaccione, *Quantifying the critical thickness of electron hybridization in spintronics materials.* Nature Communications **8,** ncomms16051 (2017).
46. G. Panaccione, M. Altarelli, A. Fondacaro, A. Georges, S. Huotari, P. Lacovig, A. Lichtenstein, P. Metcalf, G. Monaco, F. Offi, L. Paolasini, A. Poteryaev, O. Tjernberg, and M. Sacchi. *Coherent Peaks and Minimal Probing Depth in Photoemission Spectroscopy of Mott-Hubbard Systems.* Phys. Rev. Lett. **97,** 116401 (2006).
47. M. van Veenendaal, *Competition between screening channels in core-level x-ray photoemission as a probe of changes in the ground-state properties of transition-metal compounds.* Phys. Rev. B **74,** 085118 (2006).

48. M. Taguchi and G. Panaccione, *Depth-Dependence of Electron Screening, Charge Carriers and Correlation: Theory and Experiments.* in Hard X-ray Photoelectron Spectroscopy *(HAXPES)* 197–216 (Springer, Cham, 2016). doi:10.1007/978-3-319-24043-5_9
49. F. Offi, P. Torelli, M. Sacchi, P. Lacovig, A. Fondacaro, G. Paolicelli, S. Huotari, G. Monaco, C. S. Fadley, J. F. Mitchell, G. Stefani, and G. Panaccione. *Bulk electronic properties of the bilayered manganite $La_{1.2}Sr_{1.8}Mn_2O_7$ from hard-x-ray photoemission.* Phys. Rev. B **75,** 014422 (2007).
50. T. Hishida, K. Ohbayashi and T. Saitoh, *Hidden relationship between the electrical conductivity and the Mn 2p core-level photoemission spectra in $La_{1-x}Sr_xMnO_3$. Journal of Applied Physics* **113,** 043710 (2013).
51. T. Hishida, K. Ohbayashi, M. Kobata, E. Ikenaga, T. Sugiyama, K. Kobayashi, M. Okawa, and T. Saitoh. *Empirical relationship between x-ray photoemission spectra and electrical conductivity in a colossal magnetoresistive manganite $La_{1-x}Sr_xMnO_3$.* Journal of Applied Physics **113,** 233702 (2013).
52. J. Fujii, B. R. Salles, M. Sperl, S. Ueda, M. Kobata, K. Kobayashi, Y. Yamashita, P. Torelli, M. Utz, C. S. Fadley, A. X. Gray, J. Braun, H. Ebert, I. Di Marco, O. Eriksson, P. Thunstrom, G. H. Fecher, H. Stryhanyuk, E. Ikenaga, J. Minar, C. H. Back, G. van der Laan and G. Panaccione *Identifying the Electronic Character and Role of the Mn States in the Valence Band of (Ga,Mn)As.* Phys. Rev. Lett. **111,** 097201 (2013).
53. H. Tanaka, Y. Takata, K. Horiba, M. Taguchi, A. Chainani, S. Shin, D. Miwa, K. Tamasaku, Y. Nishino, T. Ishikawa, E. Ikenaga, M. Awaji, A. Takeuchi, T. Kawai and K. Kobayashi. *Electronic structure of strained $La_{0.85}Ba_{0.15}MnO_3$ thin films with room-temperature ferromagnetism investigated by hard x-ray photoemission spectroscopy.* Phys. Rev. B **73,** 094403 (2006).
54. S. Ueda, H. Tanaka, E. Ikenaga, J. J. Kim, T. Ishikawa, T. Kawai, and K. Kobayashi. *Mn 2p core-level spectra of $La_{1-x}Ba_xMnO_3$ thin films using hard x-ray photoelectron spectroscopy: Relation between electronic and magnetic states.* Phys. Rev. B **80,** 092402 (2009).
55. M. Taguchi, A. Chainani, N. Kamakura, K. Horiba, Y. Takata, M. Yabashi, K. Tamasaku, Y. Nishino, D. Miwa, T. Ishikawa, S. Shin, E. Ikenaga, T. Yokoya, K. Kobayashi, T. Mochiku, K. Hirata, and K. Motoya. *Bulk screening in core-level photoemission from Mott-Hubbard and charge-transfer systems.* Phys. Rev. B **71,** 155102 (2005).
56. M. Taguchi, A. Chainani, S. Ueda, M. Matsunami, Y. Ishida, R. Eguchi, S. Tsuda, Y. Takata, M. Yabashi, K. Tamasaku, Y. Nishino, T. Ishikawa, H. Daimon, S. Todo, H. Tanaka, M. Oura, Y. Senba, H. Ohashi, and S. Shin. *Temperature Dependence of Magnetically Active Charge Excitations in Magnetite across the Verwey Transition.* Phys. Rev. Lett. **115,** 256405 (2015).
57. G. van der Laan, *Zen and the art of dichroic photoemission.* Journal of Electron Spectroscopy and Related Phenomena **200,** 143–159 (2015).
58. B. T. Thole and G. van der Laan *Origin of spin polarization and magnetic dichroism in core-level photoemission.* Phys. Rev. Lett. **67**, 3306 (1991).
59. K. Edmonds, G. van der Laan and G. Panaccione, *Electronic structure of (Ga,Mn)As as seen by synchrotron radiation.* Semicond. Sci. Technol. **30,** 043001 (2015).
60. L.-P. Oloff, M Oura, K Rossnagel, A Chainani, M Matsunami, R Eguchi, T Kiss, Y Nakatani, T Yamaguchi, J Miyawaki, M Taguchi, K Yamagami, T Togashi, T Katayama, K Ogawa, M Yabashi and T Ishikawa *Time-resolved HAXPES at SACLA: probe and pump pulse-induced space-charge effects.* New J. Phys. **16,** 123045 (2014).
61. L.-P. Oloff, A. Chainani, M Matsunami, K. Takahashi, T. Togashi, H. Osawa, K. Hanff, A. Quer, R. Matsushita, R. Shiraishi, M. Nagashima, A. Kimura, K. Matsuishi, M. Yabashi, Y. Tanaka, G. Rossi, T. Ishikawa, K. Rossnagel and M. Oura. *Time-resolved HAXPES using a microfocused XFEL beam: From vacuum space-charge effects to intrinsic charge-carrier recombination dynamics.* Scientific Reports **6,** srep35087 (2016).
62. L.-P. Oloff, M. Oura, A. Chainani and K. Rossnagel, *Femtosecond Time-Resolved HAXPES.* in *Hard X-ray Photoelectron Spectroscopy (HAXPES)* 555–568 (Springer, Cham, 2016). doi:10.1007/978-3-319-24043-5_20
63. T. L. Cocker, L. V. Titova, S. Fourmaux, G. Holloway, H.-C. Bandulet, D. Brassard, J.-C. Kieffer, M. A. El Khakani, and F. A. Hegmann, *Phase diagram of the ultrafast photoinduced insulator-metal transition in vanadium dioxide.* Phys. Rev. B **85,** 155120 (2012).

64. J. Zhiang, X. Tan, M. Liu, S. W. Teitelbaum, K. W. Post, F. Jin, K.A. Nelson, D.A. Basov, W.Wu and R.D. Averitt. *Cooperative photoinduced metastable phase control in strained manganite films*. Nat Mater **15,** 956–960 (2016).
65. Schoenlein, R.W. et al. *New science opportunities enabled by LCLS-II X-ray lasers.* SLAC Report SLAC-R-1053 (2015).
66. T. Tschentscher, C. Bressler, J. Grünert, A. Madsen, A. P. Mancuso, M. Meyer, A. Scherz, H. Sinn and U. Zastrau. *Photon Beam Transport and Scientific Instruments at the European XFEL.* Appl. Sci. **7(6)**, 592 (2017).
67. R. Shayduk, H. Navirian, W. Leitenberger, J. Goldshteyn, I. Vrejoiu, M. Weinelt, P. Gaal, . Herzog, C. von Korff Schmising and M. Bargheer. *Nanoscale heat transport studied by high-resolution time-resolved x-ray diffraction*. New Journal of Physics, **13**, 093032 (2011).
68. Y. Zhu, J. Hoffman, C. E. Rowland, H. Park, D. A. Walko, J. W. Freeland, P. J. Ryan, R. D. Schaller, A. Bhattacharya & H. Wen. *Unconventional slowing down of electronic recovery in photoexcited charge-ordered $La_{1/3}Sr_{2/3}FeO_3$.* Nat. Comm. **9**, 1799 (2018).
69. E. Morosan, D. Natelson, A. H. Nevidomskyy, & Q. Si, *Strongly correlated materials*. Advanced Materials, **24 (36)**, 4896 (2012).
70. E. Koch, *Exchange Mechanisms*, chap. 7 in *Correlated Electrons: From Models to Materials. Modeling and Simulation*, Vol. 2, Verlag des Forschungszentrum Jülich, (2012).

# Figure Captions

**Figure 1**: **Time resolved Magneto Optical Kerr effect results. a**. Scheme of the pump-probe MOKE experimental setup. **b**. **Main panel:** Scheme of the temperature variation of the three sets of degrees of freedom. In a half-metallic system, the spin system thermalizes last, on timescales much longer than electrons and lattice. **Inset:** Schematic representation of the three sets of degrees of freedom in a magnetic solid: charges, spins and lattice. In a half-metallic system, the direct spin-electron demagnetization channel is almost suppressed due to the absence of available spin-flip processes. **c. Main panel:** Relative reflectivity variation vs. pump-probe delay measured below and above Tc= 335 K. The dashed curve represents the dynamical reflectivity trace at 400 K scaled to match the amplitude of the initial step in the curve at 200 K. The blue-shaded region highlights the variation of reflectivity induced by the DSWT, as described in the text. DSWT is absent above Tc, confirming its magnetic origin. **Inset**: Finer time-step measurement performed at room temperature, showing the sharp increase and decay of hot-carrier induced reflectivity. **d**. Normalized complex Kerr angle modulus |Θ| vs. pump probe delay, measured at 150 K and 300 K. Each curve is obtained as the modulus of the vector sum of independently recorded ellipticity and rotation traces, as sketched in the inset. While in the greyed out area the signal is still influenced by charge dynamics, a long timescale demagnetization is evident. A minimum is observed for the 150 K curve.

**Figure 2: Critical behaviour of the magnetic dynamics. Inset:** temperature dependence of Kerr ellipticity dynamics. Curves follow the colour code (corresponding to different temperatures) depicted in the bar at the right side. A minimum is reached around 250 ps for lowest temperatures. As temperature increases, the minimum shifts to larger delays moving out of the measurable interval. **Main panel:** critical slow-down modelling. Filled circles are the result of a batch fit of curves in the inset (same colour code), representing the characteristic time of ellipticity quench vs temperature. The results of the fit, performed between 10 ps and the absolute minimum of each ellipticity curve, are shown in the inset as grey solid lines for 163 K and 305 K. The vertical error bars are given by the standard deviation of the fit, hence error bars for temperatures close to Tc (vertical red line) are larger due to a reduced absolute signal. The dashed line is the fit of the power law, resulting in values of the critical exponent as described in the text.

**Figure 3. Decomposition of HAXPES spectra by screening channel a. Main panel:** Experimental HAXPES Mn 2p spectrum of half-metallic LSMO, measured at hv = 7940 eV and at T=85 K; the presence of well-screened satellite at the low binding energy side of the main peaks is indicated by the arrows. The blue line is the Tougaard background used to obtain background subtracted spectra in Fig. 4. **Inset:** intensity variation of the well-screened peak relative to Mn $2p_{3/2}$ vs. temperature is shown. **b.** Theoretical Mn 2p spectrum calculated with $Mn^{3+}$ valence (sum, blue solid line), separated in its initial state components (curves and hystograms): the ground state $3d^4$ (yellow, bottom curve), the charge transfer from the ligand hole $3d^5L$ (orange, middle) and the charge transfer from the well screened state, $3d^5C$ (red, top). The energy position of the well-screened satellites is highlighted by the grey-shaded vertical bar.

**Figure 4. Dichroism in HAXPES and theoretical analysis.** LSMO Mn 2*p* core level spectra measured at T=200 K, i.e. below Tc, with left- and right-circularly polarized X-rays, respectively labeled LCP and RCP (top spectra, triangles). Differences as measured at T=200 K (diamonds, orange) and T=300 K (squares, yellow) are shown below. At both temperatures, the largest magnetic signal is located at the energy position of the well-screened satellites. In the bottom part of the panel, calculated spectra with hybridization V* = 1.17 eV are shown (blue and red solid curves), with differences for V* = 1.17 eV (orange, solid) and V* = 0 eV (yellow, solid). Lineshape analysis confirms that hybridization must be taken into account to obtain good agreement with experimental results.

**Figure 5. TR-HAXPES lineshape variation. a.** Sketch of the pump-probe HAXPES setup, with main values of the pump-probe scheme and geometric parameters. **b.** Evolution of the Mn $2p_{3/2}$ peak in LSMO after optical pumping with 800 nm, 100 fs IR laser at a fluence of 4 mJ/cm$^2$, measured with a 50 ps synchrotron radiation probe at hv= 7940 eV varying the pump-probe delay. **c.** Variation of the lineshape of the Mn $2p_{3/2}$ peak. The topmost curve shows the difference between a static spectra measured at 150 K and at 300 K, highlighting the shifts of spectral weight observable in a quasi-static temperature variation. The other curves are obtained by subtraction of the lineshape "laser-off" from the one observed at each delay. At negative delays the absence of significant features confirms the full recovery of the initial state. At time zero, the increase of spectral weight expected from quasi-static temperature changes at high binding energy is not yet present, while the low binding energy structure is significantly reduced. On the timescales of the pulse duration (50 ps), the electronic structure is therefore in a long-lived non-equilibrium condition. At positive delays (+100 ps, +300 ps), a higher temperature transient equilibrium is formed: spectral weight changes are analogous to quasi-static lineshape variations. At long delays (1-3 ns) the differences slowly decrease.

**Figure 6. Time trace of the low binding energy structure evolution.** Each point in the graph (filled squares with error bar) corresponds to the area of the difference between the no-pump spectrum and the spectrum recorded at a fixed pump-probe delay, in the range of energies 640.5-638 eV. To calculate the relative variation, it is divided by the area of the no-pump spectrum in the same range. Spectra in the case of -100ps, +100 ps and 3.1 ns are shown in top panels, with fitted intensities of the satellites. Blue lines are guides to the eye and the vertical error bars are obtained by the integral of the difference between no-pump and the most negative delays: the absolute value of their deviation from zero gives an estimate of the uncertainty on the integrals due to the measurement noise. The horizontal uncertainty is given by the probe pulse duration of 50 ps and is smaller than the symbol size.

**Figure 1.**

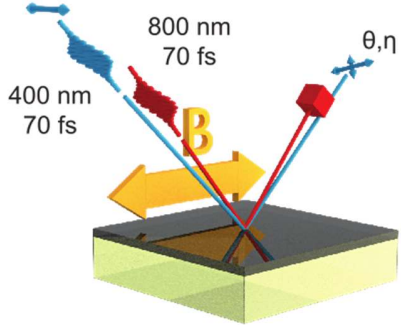
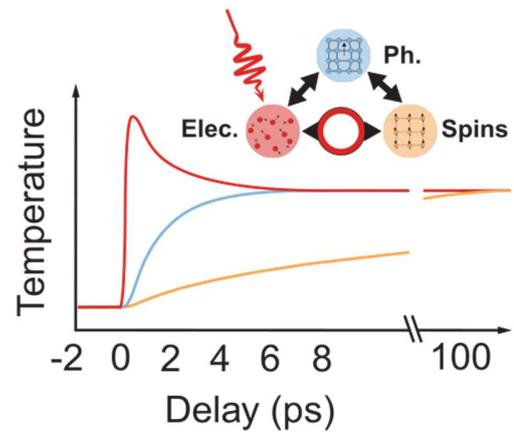
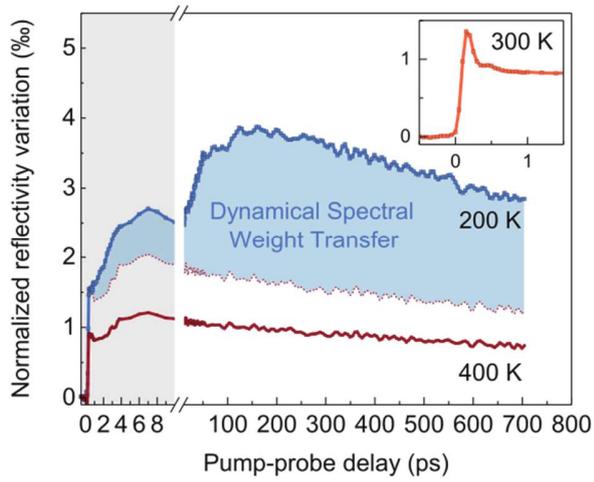
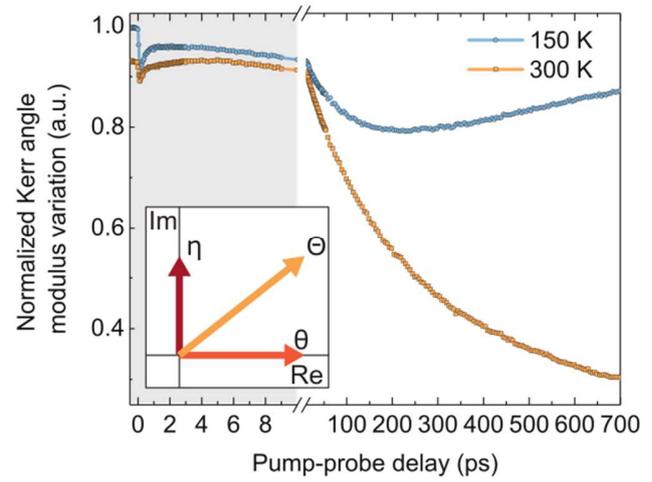

**Figure 2.**

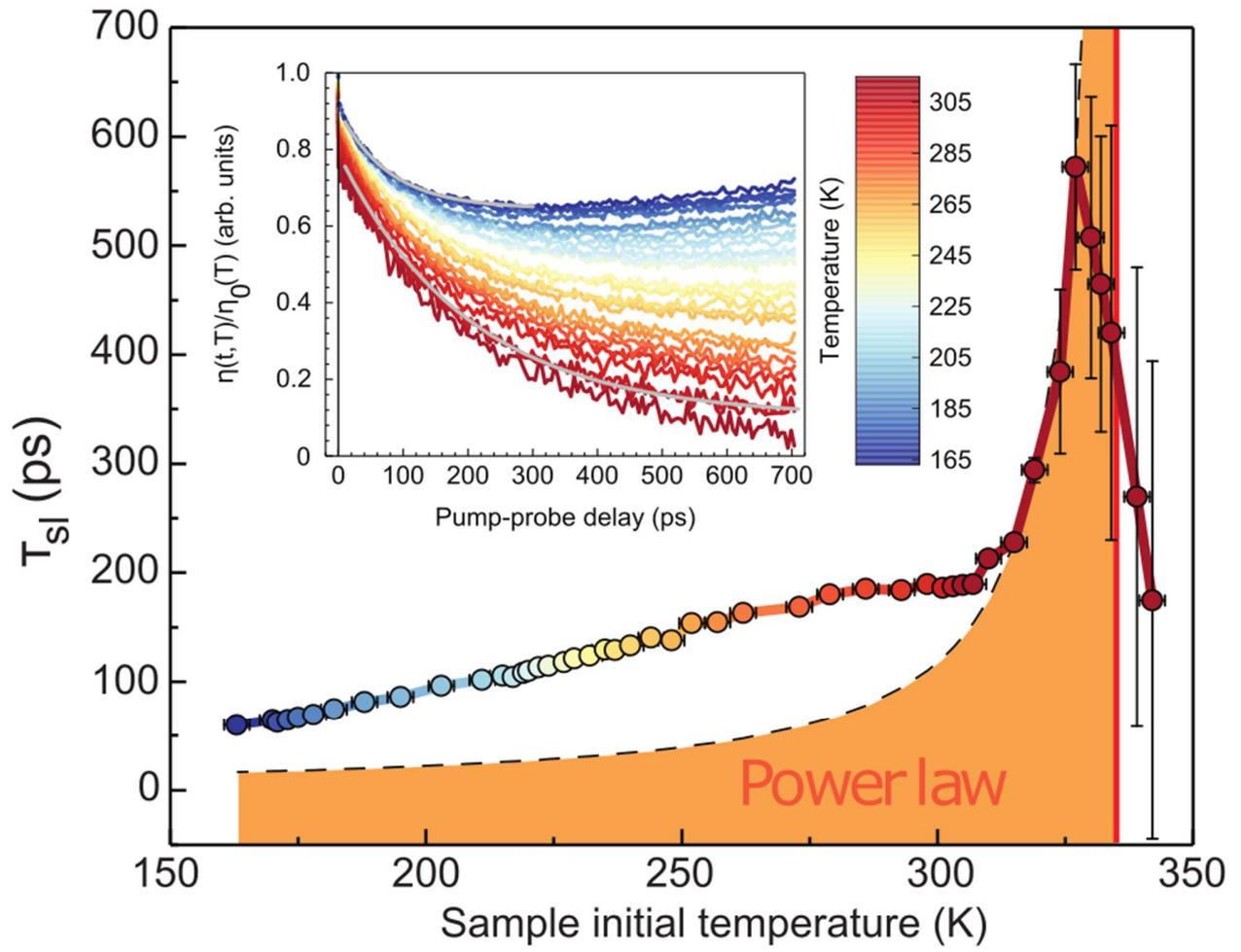

**Figure 3.**

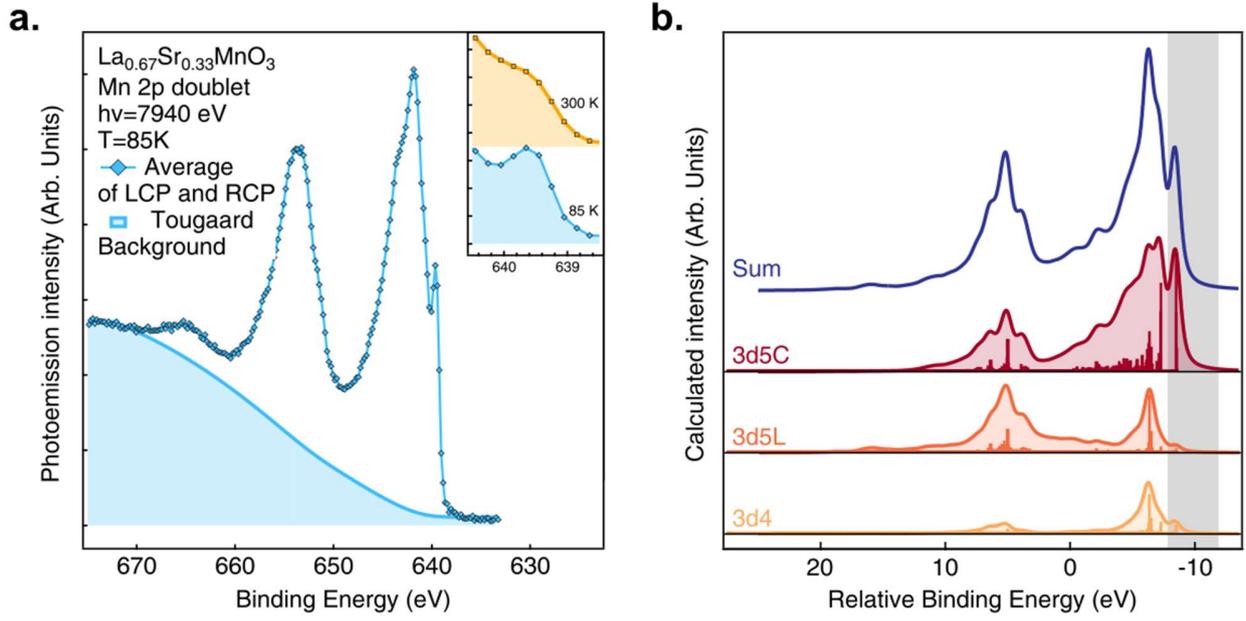

**Figure 4.**

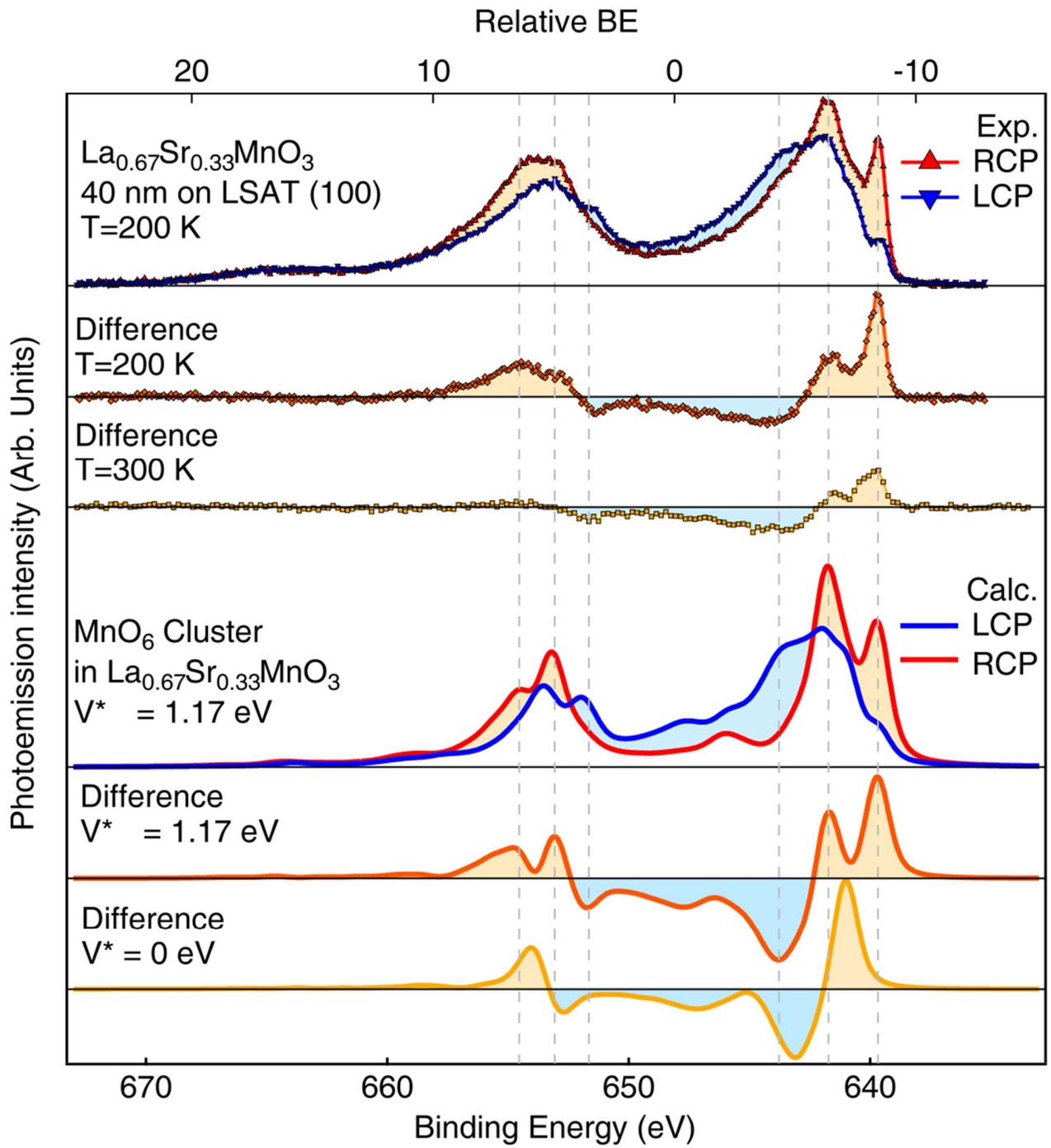

**Figure 5.**

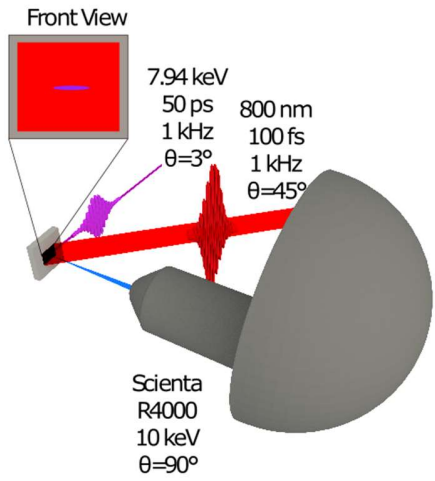
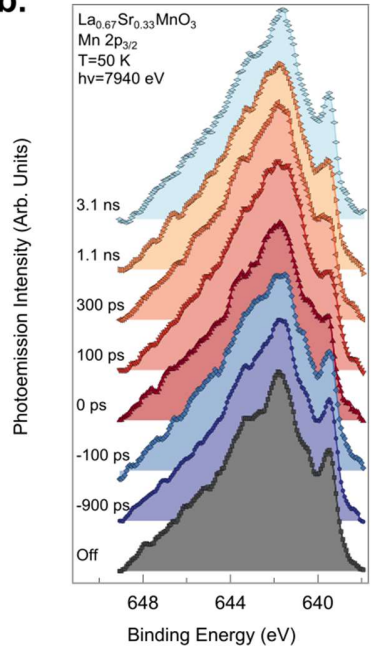
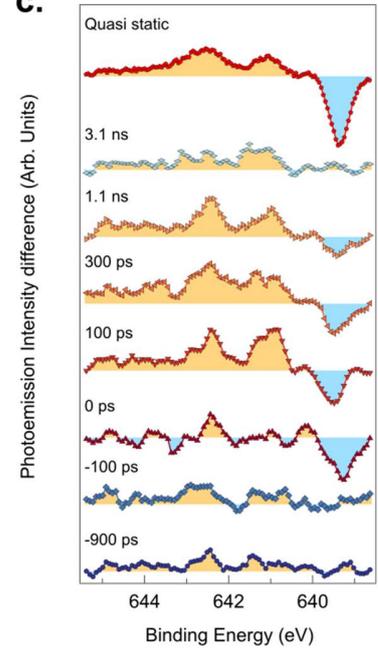

**Figure 6.**

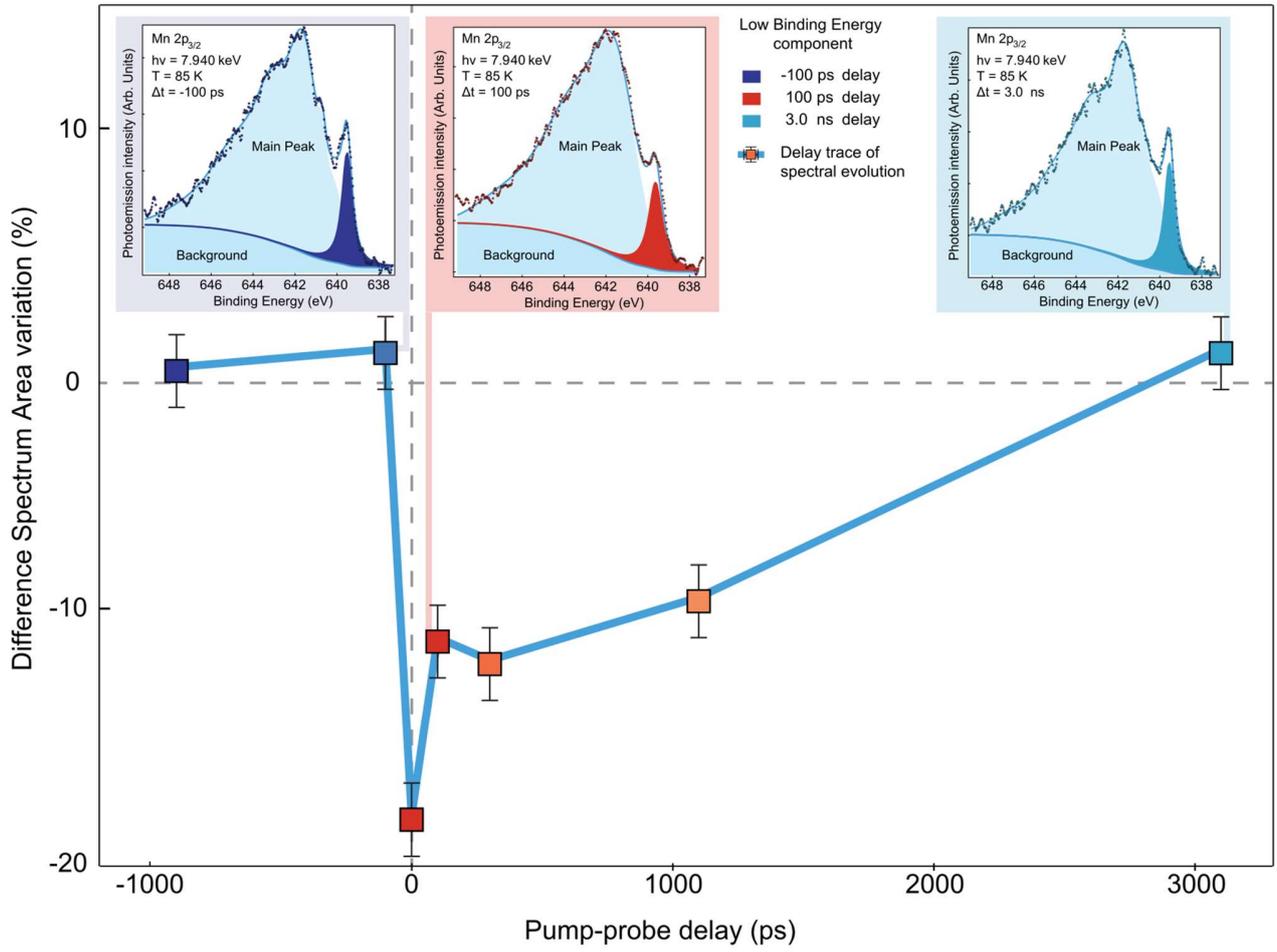